\documentclass[pre,aps,twocolumn,showpacs,preprintnumbers,superscriptaddress]{revtex4-1}

\usepackage{graphicx}
\usepackage{dcolumn}
\usepackage{bm}
\usepackage{amssymb}
\usepackage{pifont}
\usepackage{amsmath}
\usepackage{times}
\usepackage[nooneline]{subfigure}

\begin{document}

\title{Collective synchronization in the presence of reactive coupling and shear diversity}

\author{Ernest Montbri\'o}
\affiliation{Computational Neuroscience Group, Department of Information and Communication Technologies, Universitat Pompeu Fabra, 08003 Barcelona, Spain}

\author{Diego Paz\'o}
\affiliation{Instituto de F\'{i}sica de Cantabria (IFCA), CSIC-Universidad de
Cantabria, 39005 Santander, Spain }

\date{\today}

\begin{abstract}
We analyze the synchronization dynamics of a model obtained from the phase reduction
of the mean-field complex Ginzburg-Landau equation with heterogeneity. 
We present exact results that uncover the role
of dissipative and reactive couplings on the synchronization transition
when shears 
and natural frequencies are
independently distributed. As it occurs in the purely dissipative case, 
an excess of shear diversity prevents the onset of synchronization,
but this does not hold true if coupling is purely reactive.
In this case the synchronization threshold turns out to depend on the mean
of the shear distribution, but not on all the other distribution's moments.
\end{abstract}
\maketitle

Reaction-diffusion systems consisting of a large number of degrees
of freedom display a rich variety of dynamical regimes that are
important in a wide range of fields~\cite{Kur84,Mur03,CH93}.  
In particular, systems composed of many interacting aggregates of
heterogeneous, self-oscillating elements,
often show oscillations at the macroscopic level as a consequence 
of the collective synchronization of the individual oscillators
~\cite{Win80,Kur84,PRK01,Str03}.
An appropriate model to study collective synchronization is
the mean-field version of the complex Ginzburg-Landau equation
(CGLE) with heterogeneity,
\begin{eqnarray}
\label{CGLE}
\dot{z}_j &=& z_j \left[1 + i(\omega_j + q_j) - ( 1 + i q_j) |z_j|^2\right] \\
&+&\frac{K}{N} (1+ic) \sum_{k=1}^{N} (z_k- z_j). \nonumber
\end{eqnarray}
This equation describes an ensemble of $N\gg1$ globally coupled limit-cycle oscillators,
each defined by a complex variable $z_j\equiv \varrho_j e^{i\theta_j}$.
Every oscillator differs from the rest in the natural
frequency of rotation $\omega_j$ and in the shear (or nonisochronicity) $q_j$, which
measures how the frequency of rotation depends on the oscillator's
amplitude $\varrho_j$. Here we consider $\omega_j$ and $q_j$ to be independent
random variables, with a joint probability function $p(\omega, q)=g(\omega) h(q)$.

The oscillators are coupled via a diffusive coupling
of strength $K$, which has both a real (dissipative) and an imaginary 
(reactive) components. In general a positive dissipative coupling drives 
the system to a more homogeneous state \cite{MMS91} (but see \cite{DN06}).
The effect of reactive coupling on synchronization is more intricate
and strongly relies on the presence of shear $q_j\ne0$ 
\footnote{See \cite{AEK90,PRK01,Kur84} for the case of two coupled 
Stuart-Landau oscillators, and \cite{HR92,NK93,CZL+04,CRL+06} 
for the mean-field CGLE.}.

More than 30 years ago, the Kuramoto model (KM)
was proposed as an analytically tractable system to study
collective synchronization~\cite{Kur75}. Since then it has become a 
paradigmatic model to explain temporal organization in a
large variety of natural systems far from thermodynamic
equilibrium ~\cite{PRK01,ABP+05}. Under some approximations,
the KM can be rigorously obtained from Eq.~\eqref{CGLE}. Indeed,
when the mutual coupling $K$ of the oscillators is weak,
a perturbation treatment permits to reduce Eq.~\eqref{CGLE} to a set of $N$
equations for the phases only \cite{Kur84},
\begin{eqnarray}
\dot{\theta}_j  &=& \omega_j + K  (q_j - c) \label{model} \\
& + &K R  \left[(1+q_j c)\sin(\Psi-\theta_j)-(q_j-c) \cos(\Psi-\theta_j)\right],
\nonumber
\end{eqnarray}
where $R \, e^{i \Psi} = N^{-1} \sum_{k=1}^N e^{i \theta_k}$ is the complex order parameter. Originally, Kuramoto considered Eq.~\eqref{CGLE}
without reactive coupling and without shear~\cite{Kur75}. The resulting phase
equation \eqref{model} with $c=q_j=0$ is the well-known KM. Assuming constant
shear, $q_j=\hat q$, model \eqref{model} is equivalent to the
so-called Sakaguchi-Kuramoto model~\cite{SK86,NK93}. This can be seen
using the definition $\tan \beta_j=(q_j-c)/(1+q_jc)$,
with $|\beta_j|\le\frac{\pi}{2}$, which permits to write Eq.~\eqref{model}
in the more compact form
\begin{equation}
 \dot{\theta_j}  =  \omega_j + \frac{(1 + q_jc)K}{\cos\beta_j} 
\left[R \sin(\Psi-\theta_j-\beta_j) +\sin\beta_j \right]. \nonumber
\end{equation}
As in the KM, the Sakaguchi-Kuramoto model shows a transition from
incoherence to collective synchronization at large enough values of 
$K (1+\hat q c)$. The synchronized solution can be obtained 
explicitly if $g(\omega)$ is a Lorentzian distribution.

We recently reported in \cite{MP11} that, if shear is
distributed according to a certain probability function $h(q)$, the onset of
synchronization is prevented when the width of $h(q)$ exceeds a precise
threshold. These results were obtained assuming purely
dissipative coupling ($c=0$), and are fully analytic if $g(\omega)$ and
$h(q)$ are both Lorentzian.
More recently \cite{PM11} we allowed $\omega$ and $q$ to be  
non-independent, but still considering $c=0$.

Our first aim in this paper is to analyze the phase reduction \eqref{model}
with reactive coupling $(c\ne0$)
and independent random variables $\omega_j$ and $q_j$. 
We will show that in this case,  
if $g(\omega)$ and $h(q)$ are both Lorentzian, the onset of synchronization 
is also prevented beyond a critical value of the width of $h(q)$. 
In the second part we address the case of purely reactive
coupling, since it has physical relevance in the context of 
arrays of coupled nanomechanical oscillators
\cite{CZL+04,CRL+06,SIM07}, and in ion chains interacting via
Coulomb forces \cite{LC11}. We will demonstrate that, in this case, the synchronization's critical coupling becomes fully independent of the 
particular shape of the shear distribution $h(q)$. Finally we briefly 
discuss the implications of this result for the KM with
random coupling strengths recently studied by Hong and Strogatz \cite{HS11,HS11i}.
 
To analyze Eq.~\eqref{model} we adopt the thermodynamic limit $N\to \infty$.
This allows us to define a probability density function (PDF) for the phases
$f(\theta,\omega,q,t)$, such that the complex order parameter $r=R \, e^{i \Psi}$ is
\begin{equation}
r(t)=\iint_{-\infty}^{\infty} \int_0^{2\pi} e^{i \theta}
f(\theta,\omega,q,t)~ d\theta ~ d\omega ~ dq. \nonumber
\end{equation}
The evolution of Eq.~\eqref{model} obeys the continuity equation
\begin{eqnarray}
\label{continuity}
\partial_t f &=& \\
&-& \partial_\theta
\left( \left\{\omega+K(q-c) +  \tfrac{K}{2i}\left[r e^{-i \theta}C(1 - i q) - {\rm c.c.}\right] \right\} f \right). \nonumber
\end{eqnarray}
Here $C \equiv 1+ic$, and c.c.~stands for complex conjugate of the preceding term.
Next we expand $f$ in Fourier series as
$f(\theta,\omega,q,t)=\tfrac{1}{2\pi} g(\omega)h(q)
\sum_{l=-\infty}^\infty  f_l(\omega,q,t) e^{il\theta}$, 
with $f_l=f_{-l}^*$, and $f_{0}=1$.
Substituting the Fourier series
into the continuity equation
(\ref{continuity}), we obtain the infinite set of integro-differential equations
\begin{eqnarray}
\partial_t f_l =&-&i l [\omega+ K (q-c) ]  f_l  \label{fourier_set}\\
&+&\tfrac{K l}{2} \left[ r^* C^*(1+iq) f_{l-1}  -  r  C(1-iq) f_{l+1} \right] . \nonumber
\end{eqnarray}
The next step is to assume
that the asymptotic solutions of the model belong to the family of functions
\begin{equation}
f_l(\omega,q,t)=\alpha(\omega,q,t)^l,
\label{ansatz}
\end{equation}
a type of ansatz discovered by Ott and Antonsen \cite{OA08,OA09,OHA11}.
This solution of Eq.~\eqref{fourier_set} requires  $\alpha$ to evolve
according to
\begin{eqnarray}\label{alpha}
\partial_t \alpha  = &-&i[\omega+ K(q-c)] \alpha \\
&+& \tfrac{K}{2} \left[  r^* C^* (1+iq)   
 - r C (1-iq) \alpha^{2}\right], \nonumber
\end{eqnarray}
with
\begin{equation}
r^*(t)=\iint_{-\infty}^{\infty} g(\omega)h(q) \alpha(\omega,q,t) ~ d\omega ~ dq .
\label{z1}
\end{equation}

A considerable simplification is achieved if
$g(\omega)$ and $h(q)$ are chosen to be Lorentzian PDFs
\begin{equation}
g(\omega)= \frac{\delta/\pi}{(\omega-\omega_0)^2+\delta^2};~~h(q)=
\frac{\gamma/\pi}{(q-q_0)^2+\gamma^2}.
\label{lorentzians}
\end{equation}
In this case the integral \eqref{z1} can be solved closing the integrals
at infinity and using the residue's theorem;
notice that
$g(\omega)=(2\pi i)^{-1}[(\omega-\omega_0-i\delta)^{-1}-(\omega-\omega_0+i\delta)^{-1}]$,
likewise for $h(q)$.
The important requirement is that the complex function $\alpha$ can be analytically
continued from real $\omega$ and $q$ into the complex planes
$\omega=\omega_r+i\omega_i$ and $q=q_r+iq_i$, inside the integration contours.

It can be shown that $\alpha$ has no singularities in the lower half
$\omega$-plane \cite{OA08}. Regarding the variable $q$, we follow
the reasoning in \cite{MP11} and find that $\alpha$ is analytic
in the lower half complex $q$-plane for $K>0$, and in the upper one for $K<0$.
However, now this holds true only if the order parameter satisfies
\footnote{In the semicircular path at infinity ($|q|\to\infty$)
parametrized by $\vartheta$, $q=|q|e^{i\vartheta}$,
the evolution equation of $|\alpha|$ evaluated at $|\alpha|=1$ 
yields at leading order
$\partial_t |\alpha|= K(1-R |C| \cos\chi)|q| \sin\vartheta$,
where $\alpha=|\alpha|e^{-i\psi}$, $C=|C|e^{i\zeta}$,
and $\chi=\psi(q,t)-\Psi(t)-\zeta$.  Thus, the condition
$\partial_t |\alpha|< 0$   assuring that Eq.~(\ref{ansatz}) remains
finite, is fulfilled if $R< R_\times=|C|^{-1}$, in $\vartheta\in(0,\pi)$
for $K<0$, and in $\vartheta\in(-\pi,0)$ for $K>0$.}
\begin{equation}
R< R_\times=\frac{1}{\sqrt{1+c^2}}.
\label{condition}
\end{equation}
We assume that states fulfilling this condition are correctly analyzed within this framework. As we show below, the numerical simulations fully confirm 
the validity of this assumption.

\begin{figure}[t]
\centerline{\includegraphics* [width=85mm,clip=true]{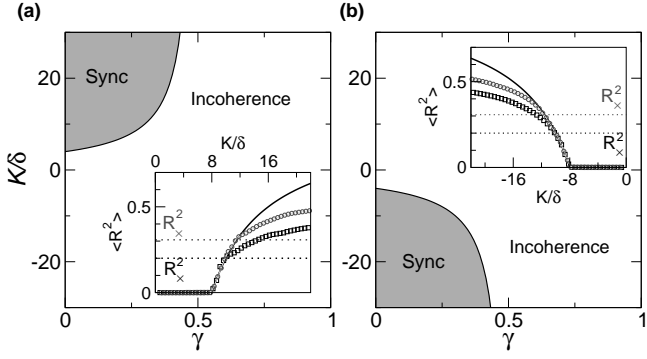}}
\caption{Phase diagrams with
boundary \eqref{bounds} for (a) $q_0c=-\tfrac{1}{2}$,
and (b) $q_0c=-\tfrac{3}{2}$. Insets show the 
numerical results of the time averaged quantity $R^2$ 
vs. $K/\delta$, with
parameters $\gamma=\tfrac{1}{4}$ ($|K_c|=8 \delta$), $\delta=0.1$,
and $\omega_0=3$.
$\{\omega_j,q_j\}_{j=1,\ldots,N}$
were deterministically selected to represent the distribution
\eqref{lorentzians}
with $N=14400$ 
oscillators. The data sets correspond to two different combinations of
$q_0$ and $c$, with $c=2$ ($R_\times^2=\tfrac{1}{5}$) and $c=\tfrac{3}{2}$
($R_\times^2=\tfrac{4}{13}$). The solid line is Eq.~\eqref{rhos}.
}
\label{fig1}
\end{figure}

Therefore, using the residue's theorem, the integrals in \eqref{z1} give
\begin{equation}
r^*(t)=\alpha(\omega=\omega^p,q=q^p,t) \equiv a(t),
\label{alpha3}
\end{equation}
where $\omega^p=\omega_0-i\delta$ and $q^p=q_0\mp i\gamma$ (for positive and negative $K$, respectively)
correspond to the simple poles of the Lorentzian PDFs \eqref{lorentzians}.
The infinite set of ordinary differential
equations \eqref{alpha} then simply reduces to the single ordinary differential
equation with complex variable
\begin{equation}
\dot a= -i\omega^p a+ \frac{K}{2} C (1-i q^p)(1-|a|^2) a. 
\nonumber
\end{equation}
As $a=R \,e^{-i\Psi}$, the equations for the order parameter inside the 
manifold defined by Eq.~\eqref{ansatz}, read
\begin{eqnarray}
 \dot R &=& \left[ -\delta + \tfrac{K}{2}(1+ c q_0  \mp \gamma)(1-R^2) \right] R,
\label{rho}\\
\dot \Psi &=& \omega_0 + \tfrac{K}{2}\left[q_0 - c (1 \mp \gamma)\right]  (1-R^2), 
\label{psi}
\end{eqnarray}
which we conjecture are the correct equations for the evolution of the order
parameter, as far as condition \eqref{condition} is fulfilled.
From Eq.~\eqref{rho} we find that a synchronized solution 
bifurcates from incoherence at the critical coupling 
\begin{equation}
K_c=
\begin{cases} \tfrac{2\delta}{1+q_0c-\gamma} & \text{if $1+q_0c>0$;}
\\
\frac{2\delta}{1+q_0c+\gamma} &\text{if $1+q_0 c< 0$,}
\end{cases}
\label{bounds}
\end{equation}
which only exists if 
\begin{equation}
\gamma < \gamma_d= |1+q_0 c|. 
\label{crit_disp}
\end{equation}
Otherwise, if $\gamma \geq \gamma_d$, incoherence becomes the only stable 
state for all $K$, see Fig.~\ref{fig1}. This result extends the one found 
in \cite{MP11} for $c=0$ to ensembles of oscillators globally coupled 
via \emph{both} dissipative and reactive coupling. However, as it is depicted 
in Fig.~\ref{fig1}(a), now the region of stable synchronization is located at 
positive values of $K$ only if
\begin{equation}
1 +q_0 c > 0.
\label{BF}
\end{equation}
This inequality is the well-known Benjamin-Feir-Newell criterion for the
stability of plane waves in the homogeneous
CGLE~\cite{HR92,NK93,Kur84,AK02,CH93}, which
is valid in any dimension (infinite in the present case).
Finally, we also find that the order parameter of
the partially synchronized solution follows
\begin{equation}
R^2= \frac{K-K_c}{K}, \, \mbox{with  $R<R_\times$},
\label{rhos}
\end{equation}
exactly as in the KM with Lorentzian $g(\omega)$ \cite{Kur84}. However,
now this formula holds only up to $R_\times$, recall Eq.~\eqref{condition}.
The insets in Fig.~\ref{fig1} show numerical simulations that confirm the validity
of Eq.~\eqref{rhos}. Remarkably, our numerical simulations indicate
that $R$ departs from Eq.~\eqref{rhos} precisely when $R$ exceeds $R_\times$.

Unfortunately, there is no straightforward theoretical extension
of these results to more general distributions $g(\omega)$ and $h(q)$,
but still some reasonable conjectures can be raised:
\begin{enumerate}
\item[(i)] The flip of the synchronization region when $1+q_0c$ changes sign---compare
Figs.~{\ref{fig1}}(a) and 1(b)---is a consequence of the Benjamin-Feir-Newell
criterion~\eqref{BF} and it can be supposed to be a general feature of
Eq.~\eqref{model}. 
\item[(ii)] Although it seems difficult to prove, the divergence of 
$K_c$ at a critical value of the shear diversity [Eq.~\eqref{crit_disp}] 
is likely to be a general property, as it happens in the purely 
dissipative case ($c=0$) \cite{MP11}. 
\item[(iii)] For certain distributions and parameter
values, and as a consequence of the persistence of a fully synchronized
solution $\theta_j=\Psi$ located at $|K|g(\omega_0)\to\infty$, stable
synchronization and incoherence should  coexist at large $K$ values
(as it occurs for $c=0$ and Gaussian distributions \cite{MP11}). Note that
infinitesimal perturbations obey $\dot{\delta\theta}_j=\tfrac{K}{N}(1+q_jc)
[(1-N)\delta\theta_j+\sum_{l\ne j}\delta\theta_l]$. The Jacobian matrix has
always one  trivial zero eigenvalue.
If $c=0$ and $K$ positive, the remaining eigenvalues are negative, and the
fixed point is stable irrespective of the width of $h(q)$. However, if
$c\ne0$ the fixed point becomes a {\em saddle} when the $q_j$'s exceed
some degree of heterogeneity, and hence its continuation at finite $K$ is
not an attractor either. In sum, under a large enough
heterogeneity of shear, incoherence should be the only attractor at all $K$ values;
however, if $c=0$, synchronization coexists with incoherence at large enough $K$---provided $h(q)$ is not heavy-tailed, see \cite{MP11}.
\end{enumerate}

For the remainder of this article, we will concentrate on the case
of {\em purely reactive coupling}. Motivated by the
dynamics of nanoscale mechanical oscillator arrays, this problem
was analyzed in detail by Cross et al.~\cite{CZL+04,CRL+06}
with a coupling of the form $i \tfrac{\kappa}{N} \sum_k (z_k-z_j)$
in Eq.~\eqref{CGLE}, and without shear diversity.
To investigate the effect of shear diversity, we first write
the phase model~\eqref{model} without dissipative coupling. Substituting
$c=\kappa/K$ in Eq.~\eqref{model} and letting $K\to0$ yields
\begin{equation}
\label{modelr}
\dot{\theta}_j=\omega_j -\kappa + \kappa R  \left[ q_j \sin(\Psi-\theta_j)+\cos(\Psi-\theta_j)\right],
\end{equation}
where $\kappa$ is now the total reactive coupling. Then, Eq.~\eqref{bounds}
suggests that in this limit the critical coupling is
\begin{equation}
\kappa_c=\frac{2\delta}{q_0},
\label{kappac_lor}
\end{equation}
that remarkably depends on $q_0$ but
is independent of the amount of heterogeneity $\gamma$.

The derivation of Eq.~\eqref{kappac_lor} is not rigorous because $R_\times=0$
in this limit, and condition \eqref{condition} is not fulfilled.
Therefore, to confirm the validity of Eq.~\eqref{kappac_lor} and to determine 
how this result generalizes to other distributions, next we
perform the linear stability analysis of the incoherent state of
Eq.~\eqref{modelr} \cite{SM91}. In the incoherent state all modes $f_l$,
save the trivial one $f_0=1$, vanish. The equation for the Fourier modes,
related to Eq.~\eqref{fourier_set}, is
\begin{equation}
\partial_t f_l =-i l (\omega-\kappa)  f_l   \nonumber
+\frac{\kappa l}{2} \left[ r^* (q-i) f_{l-1}  -  r  (q+i) f_{l+1} \right] . \nonumber
\end{equation}

Linearizing this equation about the incoherent state, we find that the only
potentially unstable mode is the $l=1$
\begin{eqnarray}
\partial_t f_1=&-&i(\omega-\kappa) f_1 \nonumber\\
&+& \frac{\kappa}{2}(q-i) \iint_{-\infty}^\infty f_1(\omega',q',t) g(\omega') h(q') d\omega' \,dq'.\nonumber
 \end{eqnarray}
Let $f_1(\omega,q,t)=b(\omega,q) e^{\lambda t}$, and neglect the trivial solution
$b=0$. Invoking self-consistency and separating $\lambda$ into its real and imaginary parts
($\lambda=\lambda_r+i\lambda_i$) yields
\begin{equation}
 \frac{2}{\kappa}=\iint_{-\infty}^\infty \frac{(q-i)[\lambda_r-i(\omega-\kappa+\lambda_i)]}
{\lambda_r^2+(\omega-\kappa+\lambda_i)^2} g(\omega) h(q) d\omega \, dq. \nonumber
\label{interesting}
\end{equation}
The interesting feature in the right hand side of this equation
is that the integration over $q$
is trivial and the result does \emph{not} depend on the particular shape of $h(q)$.
Performing the limit $\lambda_r\to0^+$ to obtain the stability threshold
$\kappa_c$ yields
\begin{equation}
 \frac{2}{\kappa_c}= (q_0-i)\left[\pi g(\kappa_c-\lambda_i)
-i\int_{-\infty}^\infty \frac{g(\omega)}{\omega-\kappa_c+\lambda_i} d\omega\right],
\label{kappac}
\end{equation}
that only depends on $h(q)$ through its mean value $q_0$ (defined as principal
value if required). Finally, splitting \eqref{kappac} into its real and imaginary
parts we obtain a system of two equations for the unknowns $\kappa_c$ and $\lambda_i$
\begin{eqnarray}
(1+q_0^2)\pi g(\kappa_c-\lambda_i)=\frac{2q_0}{\kappa_c}, \label{s1}\\
(1+q_0^2) \int_{-\infty}^\infty \frac{g(\omega)}{\omega-\kappa_c+\lambda_i} d\omega=-\frac{2}{\kappa_c}.  \label{s2}
\end{eqnarray}
These equations can be solved for Lorentzian $g(\omega)$, and indeed we
recover the boundary \eqref{kappac_lor}. However, note that now this result is
stronger, since we have not imposed any constraint
on the shape of $h(q)$. Here $h(q)$ can be \emph{any} distribution of mean $q_0$.  
Additionally, an explicit value for $\kappa_c$ can be easily obtained from Eq.~\eqref{s1}
if $g(\omega)$ is a uniform (top-hat) distribution.
These results for Lorentzian and uniform $g(\omega)$ 
are in agreement with those obtained in \cite{CRL+06} with $h(q)=\delta(q-q_0)$.
This confirms that the phase equation is indeed a
good approximation of the amplitude equation in the limit of 
weak coupling and narrow frequency distributions.

\begin{figure}
\centerline{\includegraphics* [width=85mm,clip=true]{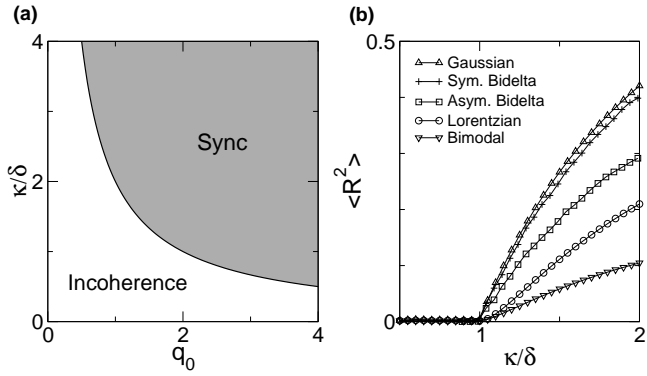}}
\caption{(a) Phase diagram with boundary \eqref{kappac_lor}
in the case of purely reactive coupling and $g(\omega)$ Lorentzian.
(b) Time averages $\left<R^2\right>$ obtained from numerical 
simulations with $q_0=2$,
$\delta=1$, $N=40000$, and for different distributions of $q$:
Gaussian with variance $\nu^2=1$,
symmetric bidelta: $h(q)=\tfrac{1}{2}\delta(q-3)+\tfrac{1}{2}\delta(q-1)$,
asymmetric bidelta: $h(q)=\tfrac{3}{4}\delta(q-3)+\tfrac{1}{4}\delta(q+1)$,
Lorentzian with $\gamma=1$, and bimodal: $h(q)=
\tfrac{1/(2\pi)}{[q-(q_0-5/2)]^2+1}+
\tfrac{1/(2\pi)}{[q-(q_0+5/2)]^2+1}$. 
In all cases the critical coupling is
at $\kappa_c / \delta=1$, as predicted
by Eq.~\eqref{kappac_lor}.
}
\label{fig2}
\end{figure}

Figure \ref{fig2}(a) displays a phase diagram with the boundary \eqref{kappac_lor},
and Fig.~\ref{fig2}(b) shows the time average $\left<R^2\right>$ obtained
from numerical simulations using different distributions $h(q)$ with common $q_0$
values. As expected, in Fig.~2(b) the transition between incoherence and 
synchronization occurs at the same value of $\kappa/\delta$
irrespective of the distribution type.

Finally we point out an interesting similarity between Eq.~\eqref{modelr} 
and the model recently studied by Hong and Strogatz \cite{HS11,HS11i}, 
which in our notation reads: $\dot\theta_j=\omega_j+q_j R\sin(\Psi-\theta_j)$. 
Note that here $q_j$ acts as a distributed coupling strength. Performing
a stability analysis like we did above, we obtain that the stability border 
of incoherence satisfies $2= \pi q_0  g(\omega_0),$
if $g(\omega)$ is unimodal and symmetric. Again, we obtain a formula
that depends on the mean of $h(q)$, but not on its shape.
This result reproduces the classical Kuramoto relation
for uniform all-to-all coupling [$h(q)=\delta(q-q_0)$], and
the critical point found in Eq.~(12) of \cite{HS11} for
Lorentzian $g(\omega)$ and bidelta $h(q)$.

In conclusion, we have reported on exact results that extend the phase models
of Kuramoto and Sakaguchi \cite{Kur75,SK86} to situations where shear is distributed.
In contrast to the recent work \cite{MP11},
here the coupling is not
purely dissipative but also contains a reactive component $c$. In this case we 
also find that shear diversity prevents the onset of collective synchronization, 
but the Benjamin-Feir-Newell criterion determines now if the region of 
synchronization is located at positive or negative values of $K$.
Finally, we have obtained a remarkable result when the coupling is purely reactive:
the stability threshold of incoherence depends on the mean shear $q_0$, while
the shear diversity becomes irrelevant.

Financial support from the Ministerio de Ciencia e Innovaci\'on (Spain) under
Projects No.~FIS2009-12964-C05-05 and No.~SAF2010-16085 is acknowledged.


\bibliographystyle{prsty}
\end{document}